\documentclass{aa}
\date{}
\usepackage{graphicx}
\usepackage{natbib}
\usepackage{txfonts}
\def\new#1 {{\bf #1 }}
\def\cut#1 {\sout{#1} }
\def\beq{\begin{equation}}
\def\eneq{\end{equation}}
\def\simgt{\lower.5ex\hbox{$\; \buildrel > \over \sim \;$}}
\def\simlt{\lower.5ex\hbox{$\; \buildrel < \over \sim \;$}}
\begin{document}
\title{Magnetic field upper limits for jet formation}

\author{M. Massi and M. Kaufman Bernad\'o\thanks{Humboldt Research Fellow}}

\institute{Max-Planck-Institut f\"ur Radioastronomie, Auf dem H\"ugel 69, D-53121 Bonn, Germany}

\offprints{M. Massi, \\ \email{mmassi@mpifr-bonn.mpg.de}}

\abstract
{Very high magnetic fields at the surface of neutron stars or in the accretion disk of black holes inhibit the production of jets.}
{We quantify here the magnetic field strength for jet formation.}
{By using the Alfv\'en Radius, $R_{\rm A}$,
we study what we call {\it the basic condition}, $R_{\rm A}/ R_*  = 1$ or $R_{\rm A}/ R_{\rm LSO} = 1$  (LSO, last stable orbit),
 in its dependency on the magnetic field strength and the mass accretion rate, and we analyse these results in  3-D and 2-D plots in the case of neutron star and black hole accretor systems, respectively. For this purpose, we did a systematic search of all available observational data for magnetic field strength and the mass accretion rate.}
{The association of a classical X-ray pulsar (i.e. $B \sim 10^{12}$ G) with jets is excluded even if accreting
 at the Eddington critical rate. 
Z-sources may develop jets for $B \simlt 10^{8.2}$ G,  whereas 
 Atoll-sources  are potential sources of jets if $B \simlt 10^{7.7}$ G. 
It is not ruled out that a millisecond X-ray pulsar could develop jets, at least for those sources
where  B$ \simlt 10^{7.5}$ G. In this case
the millisecond X-ray pulsar could
 switch to a microquasar phase during its maximum accretion rate.
For stellar-mass black hole X-ray binaries, the condition is that $B \simlt 1.35 \times 10^8$ G and $B \simlt 5 \times 10^8$ G at the last stable orbit for a Schwarzschild and a Kerr black hole, respectively. For active galactic nuclei (AGNs), it reaches $B \simlt 10^{5.9}$ G for each kind of black hole. These  theoretical results are in complete agreement with available observational data. 
}
{}

\titlerunning{Magnetic field upper limits}

\authorrunning{M. Massi and M. Kaufman Bernad\'o}

\maketitle

\section{Introduction} \label{introduction}

\begin{figure}[h!]
\centering
\resizebox{\hsize}{!}{\includegraphics[ angle=-90, scale=1.0]{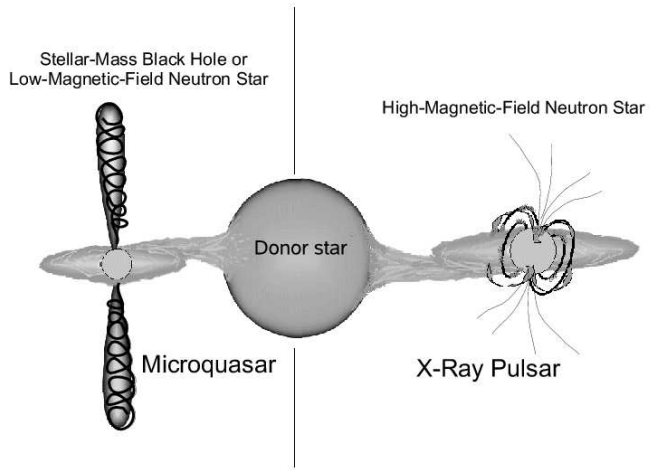}}
\caption{Left - The basic components of a microquasar: an accreting compact object (a neutron star or a stellar-mass black hole), a donor star, and radio emitting  relativistic jets.
Right - The basic components of a classical (slow) X-ray pulsar:
an accreting neutron star with a very strong magnetic field ($\ge 10^ {12} $G)
emitting X-rays at the polar caps. A millisecond X-ray pulsar ($B \simlt 10^{8}$ G) could switch between a microquasar and an X-ray pulsar scenario, depending on the accretion mass rate (see Section \ref{millisec}). Figure not drawn to scale.}
\label{micro+pulsar}
\end{figure}

As its name suggests, a microquasar  is a miniature version of a quasar:
an accretion disk with a radius of a few thousand kilometers surrounds a compact object 
-a black hole (BH) of a few solar masses or a neutron star (NS)- that accretes from a companion star
and two relativistic jets that are propelled out of the disk (Fig.~\ref{micro+pulsar})
by the same processes as in a quasar.

Because microquasars are present in our own Galaxy,  
the  study of the  evolution of their relativistic jets only requires 
a few days rather than the years necessary 
to measure appreciable proper motions for the radio jets
of far-away quasars.
Also, when looking at the intrinsic variability
of microquasars, these comparatively ``small'' objects change faster than quasars:
considering $\tau\sim R_{\rm Schwarzschild}/c\propto M_{\bullet}$ as a characteristic time scale for variations,
phenomena with timescales of minutes in microquasars
with BHs of 10 $\rm M_{\odot}$ would take years in AGNs with supermassive BHs of $10^7$ $\rm M_{\odot}$.
This  enormous difference provoked strong interest in microquasars, 
leading to this field of astrophysics developing very quickly in the past decade.

Recently another aspect of microquasars has begun to be considered. The presence of BHs or NSs as the accreting object could
lead to a better understanding of the mechanisms of jet production.
The most promising process producing relativistic jets
involves magneto-hydrodynamic centrifugal acceleration of material 
from the accretion disk. Does the ergosphere of a Kerr black hole also play a role in the jet-acceleration?
The existence of  microquasars with different ``engines"
may help to distinguish  better between the relative importance of
magnetic field and disk rotation with respect to the
ergosphere  (Meier et al. 2001). This last ``ingredient" is obviously missing in a neutron star.

Microquasars are a subclass of the stellar systems called X-ray binaries (XRBs). These systems are formed by two stars
of very different nature: a normal star acting as a mass donor
and a compact object, the accretor, that can be either a NS or a BH. XRBs are classified into low-mass X-ray binaries (LMXBs) and
high-mass X-ray Binaries (HMXBs) depending on the mass of 
the companion star (van Paradijs \& McClintock 1996).
HMXBs have young bright stars (O-B) and LMXBs instead
have  old stars (later than G).
This classification leaves the nature of the accreting object unspecified. 

Microquasars are defined as the XRB systems where either high-resolution radio interferometric techniques have shown the presence of collimated jets (Table~\ref{microquasars}) or a flat/inverted radio spectrum has been observed (indirect evidence of an expanding  continuous jet, e.g. XTE J1118+480, XTE J1859+226, V404 Cyg, GRO J0422+32, see Fender 2001, 2004). The nature of the compact object, NS or BH, is still uncertain for several microquasars (see Table~\ref{microquasars}). 

When will an accreting
neutron star become a microquasar and  when, on the other hand, an X-ray pulsar? When will a BH XRB system be able to  evolve into a microquasar phase? We  analyse here the {\it initial conditions} for an ejection event to be possible in such systems. A low magnetic field at the NS surface or at the last stable orbit of the accretion disk will be a necessary initial condition. We aim to quantify this important parameter here and therefore give an upper limit for the magnetic field strength for which an ejection could happen in a NS or BH XRB system.
We  also predict the corresponding behaviour for AGNs  
using standard scaling
(Merloni et al. 2003, Falcke et al. 2004, McHardy et al. 2006).
 
\section{Magnetohydrodynamic jet production}\label{MHD}

\begin{table}
\begin{center}
\caption[]{\label{microquasars}XRB systems with resolved radio jets.}
\begin{tabular}{llll}
\hline \hline \noalign{\smallskip}
Name& Companion & Accretor& Jet size (AU)  \\
\noalign{\smallskip} \hline \noalign{\smallskip}
&~~~~~~~~~~HMXBs&&\\
\noalign{\smallskip} \hline \noalign{\smallskip}
LS I +61 303& B0V              & NS/BH ?& $10-700$       \\
V 4641 Sgr  & B9III & Black Hole & $-$ \\
LS 5039  & O6.5V((f))              & NS/BH ? & $10-1000$              \\
SS 433        &  evolved A      &   NS/BH ?   & 10$^4-10^6$ \\
Cygnus    X-1    & O9.7Iab& Black Hole    &40  \\
Cygnus    X-3    & WNe&   NS/BH ?  & $10^4$ \\
\noalign{\smallskip} \hline \noalign{\smallskip}
&                    ~~~~~~~~~LMXBs&&\\
\noalign{\smallskip} \hline \noalign{\smallskip}
Circinus X-1 & Subgiant & Neutron Star    &$10^4$   \\
XTE J1550-564 & G8-K5V &Black Hole   &$10^3$     \\
Scorpius X-1 & Subgiant & Neutron Star    &40  \\
GRO J1655-40 & F3/5IV &Black Hole       &8000 \\
GRS 1915+105  & K-M III & Black Hole & $10-10^4$      \\
GX 339-4  &  &Black Hole     &$< 4000$   \\
1E 1740.7-2942& & NS/BH ? &$10^6$\\
XTE J1748-288  &  &  NS/BH ?  &$10^4$    \\
GRS 1758-258  & &  NS/BH ?    &$10^6$  \\
\noalign{\smallskip} \hline \hline
\end{tabular}
\end{center}
{\small Massi 2005; Paredes 2005; Casares 2005}
\end{table}

Numerical simulations show that the launch of a jet involves a weak 
large-scale poloidal magnetic field anchored in rapidly 
rotating disks or compact objects  (Meier et al. 2001).
The geometry of this field is analogous to the one present in solar coronal holes,
and it could be  generated by a dynamo process (Blackman \& Tan 2004).

The strength of the large-scale poloidal field must be low enough
for the  plasma pressure $P_{\rm p}$, to  dominate the magnetic field pressure $P_{\rm B}$ (Blandford 1976). 
Only under that  condition, $P_{\rm B} < P_{\rm p}$, is the differentially rotating  disk  able to bend
the magnetic field lines in a magnetic spiral (Meier et al. 2001).
Because of  the increasing  compression of the magnetic field lines,
the magnetic pressure will grow and may  become higher 
than the gas pressure on the surface of the accretion disk, where the density is lower (see the role of the thick disk in Meier 2001). 
There, the magnetic field  becomes ``active", i.e. dynamically
dominant, and  the plasma has to follow the twisted magnetic field lines, 
creating  two spinning-plasma flows.

\begin{figure}[t!]
\centering
\resizebox{\hsize}{!}{\includegraphics[angle=0, scale=0.75]{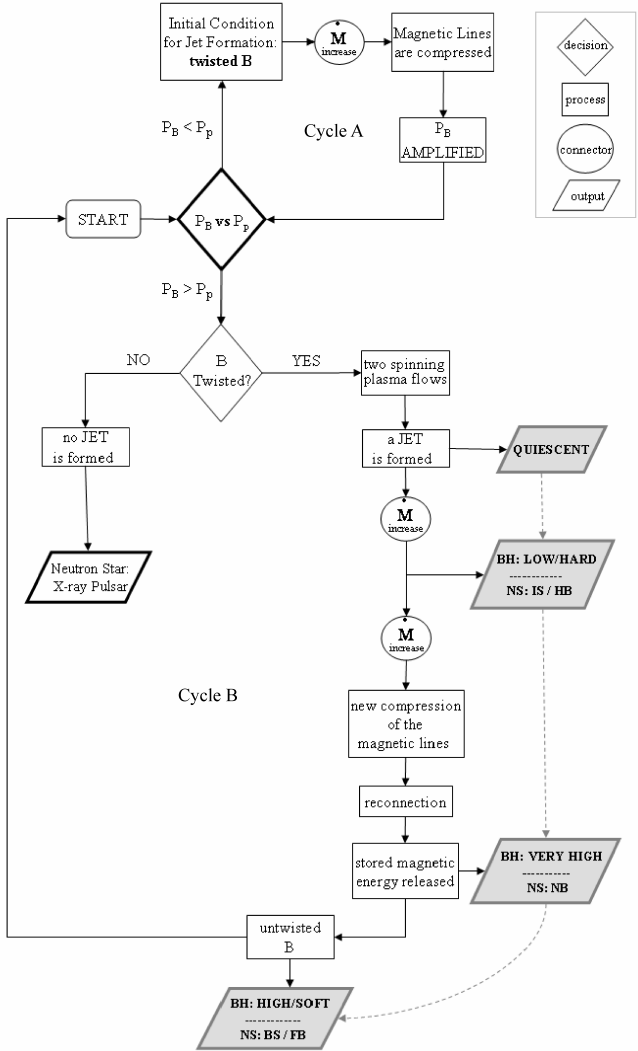}}
\caption{Flowchart for the jet formation process. We show here the parallelism between the presence of a jet and the X-ray states cycle.  
The X-ray state names are mentioned for both, BH and NS XRBs. In the case of neutron stars the distinction between atoll- and Z-type sources (see Section~\ref{AandZ}) is made: IS = Island State / HB = Horizontal Branch, NB = Normal Branch,  BS = Banana State / FB = Flaring Branch.}
\label{chart}
\end{figure}

The competition process between the magnetic field pressure and the plasma pressure that seems to be at the base for the formation of a jet has been summarised in a flowchart in Fig.~\ref{chart}. The generation of jets and their presence in XRBs is coupled to the evolution of a cycle that can be observed in the X-ray states of this kind of systems (Fender et al. 2004a, Ferreira et al. 2006)\footnote{Although the different X-ray states  receive different names depending on whether the compact object is a NS or a BH, it has been shown by Migliari \& Fender (2006) that
 the two kinds of systems parallel each other perfectly.}. We therefore  complement the jet formation flowchart showing the parallelism between the presence of a jet and the different X-ray states. The system enters in ``cycle B" (i.e. the cycle along the X-ray states) only when the condition $P_{\rm B} > P_{\rm p}$ is finally reached in ``cycle A" (see Fig.~\ref{chart}).

As recently proved for the bipolar outflows from young stellar objects,
these rotating plasma-flows take angular momentum 
away from the  disk (magnetic braking):
the angular momentum transport rate of the jet can be two thirds or more of the estimated transport rate through the relevant portion of the disk (Woitas et al. 2005).
This loss of  angular momentum slows down the disk material to sub-Keplerian rotation and therefore the
disk matter can finally accrete onto the central object (Matsumoto et al. 1996).
This increase in accreted matter implies that the material pulls 
the deformed magnetic field with it even further. The magnetic field
compression is thus increased and magnetic reconnection may occur (Novikov \& Thorne 1973, Matsumoto et al. 1996, Gouveia dal Pino 2005).
The stored magnetic energy is released and the field returns to the
state of minimum energy (i.e. untwisted).

Considering that the jet generation is conditioned by the  competition between  $P_{\rm B}$ and   $P_{\rm p}$, we can conclude that microquasars are in fact a possible phase that any low magnetic field XRB system may undergo for a certain period of time. 

Besides the cycle connected with the microquasar phase, one has to consider the situation in which $P_{\rm B} > P_{\rm p}$ from the very beginning, i.e. with an untwisted field.
This is the case of an XRB holding a NS with a
strong magnetic field. This situation corresponds to a classical X-ray pulsar scenario (see Section~\ref{classic}). The strong magnetic field cannot be  twisted. It dominates the dynamic all the time, confining 
the disk-material down to the magnetic poles where two emitting caps are created (see Fig.~\ref{micro+pulsar}). 

We  quantify these different situations in the following sections by setting an upper limit to the magnetic field strength for which the initial condition for jet formation will be fulfilled. 

\subsection{The Alfv\'en radius: a tool for the basic condition}
\label{alfven}

The magnetic field  can  be bent in a sweeping spiral only
if the  magnetic pressure, $P_{\rm B}=B^2/{8 \pi}$, is lower than
the hydrodynamic pressure, $ P_{\rm p}=\rho$v$^2$, of the accreting material. The distance at which the magnetic and plasma pressure balance each other is called the Alfv\'en radius. Using this magnitude, we define what we call {\itshape the basic condition} for jet formation, $R_{\rm A}/R_* = 1$ or $R_{\rm A}/R_{\rm LSO} = 1$, in the case of NS or BH XRBs, respectively, with $R_*$ the NS surface radius, and $R_{\rm LSO}$ the radius of the BH last stable orbit. Imposing this condition   guarantees that $P_{\rm B} < P_{\rm p}$ is valid  over the whole disk. The relation  $R_{\rm A}/R_* = 1$ or $R_{\rm A}/R_{\rm LSO}=1$ ensures that the magnetic field lines will then be twisted close to the compact object.

The expressions $R_{\rm A}/R_*$ and $R_{\rm A}/R_{\rm LSO}$ are functions of the magnetic field strength and of the mass accretion rate and will therefore allow us to establish under which combination of these parameters NS and BH XRBs may undergo a microquasar phase.

\section{Neutron star X-ray binaries}\label{NSXRB}

By equating $B^2/{8 \pi}$ and $\rho$v$^2$, one can get an expression for $R_{\rm A}$. In accreting neutron stars, there is  the unknown magnetic field of the disk and the stellar magnetic field for which observations exist. The assumption that the stellar field dominates the disk field allows a quantitative estimate.

Expressing the mass accretion rate $\dot{M}$ as $4 \pi R^2 \rho \rm v$ (Longair 1994), where v is the infall velocity 
$\rm{v}=(2 G M_* / R) ^{1/2}$ and for a magnetic dipole field with a surface magnetic field  $B_*$,
${B/ B_*}=[{R_* / R}]^{3}$, we get

\beq
R_{\rm A}= { B_*}^{4/7} {R_*}^{12/7}\left ({2GM_*} \right)^{-1/7}{\dot M}^{-2/7}.
\label{ar}
\eneq

\noindent
The ratio $R_{\rm A}/R_*$  in terms of the accretion rate, $\dot{M}$, 
and the NS surface magnetic field, $B_*$, is therefore equal to 

\beq
R_{\rm A}/R_* \simeq  0.87 \left ( { B_*\over 10^8 ~ \rm G} \right)^{4/7} \left ( {\dot M\over {10^{-8}~ {\rm{M}_{\odot}\over yr}}} \right)^{-2/7},
\label{basic}
\eneq

\noindent for a neutron star with a mass and a radius of $M_*=1.44~ \rm{M}_{\odot}$ and $R_*= 9$ km 
(Titarchuk \& Shaposhnikov  2002).

Table~\ref{mdot} shows the  ranges, available in the literature, for accretion rate and magnetic field strength of NS in  XRB systems.
Including classical X-ray pulsars,
the  interval for $B$  ranges over more than 4 orders of magnitude:
from classical X-ray pulsars with fields above 10$^{12}$G, 
to the low  value of 10$^{7-8}$G for the other sources. 
The interval for  accretion rate covers several orders of
magnitude as well, from less than 0.1\% of the Eddington critical rate (see Eq.~\ref{accedd} in Section \ref{BHXRB})
for  millisecond X-ray pulsars to  Eddington critical rate 
for the  Z sources (see references in Table~\ref{mdot}).

Inserting the values of Table~\ref{mdot} into Eq.~\ref{basic}, we obtain a 3-D plot of the parameter $R_A/R_*$ as function
of both the accretion rate 
and the magnetic field strength, which we show in Fig~\ref{3-D}. The ``white area" refers to values of $R_A/R_* = 1$. This is the region
where the accretion rate and the magnetic field strength are combined in such a way
that the stellar field is not dynamically important at any point; therefore, this white region corresponds to the range of values in the parameter space where potential microquasars exist. One can see in Fig.~\ref{3-D} that this region is rather  small for the given wide  range of $B$ and $\dot M$. 

\begin{table}[t!]
\begin{center}
\caption[]{\label{mdot} Neutron stars: Accretion mass rate and magnetic field strength}
\begin{tabular}{ccc}
\hline \hline \noalign{\smallskip}
Class& ${\dot M\over 10^{-8}}  (\rm{M}_{\odot}~\rm{yr}^{-1})$& $B_*$ (G) \\
\noalign{\smallskip} \hline \noalign{\smallskip}
classical X-ray& 0.025 - 0.1 (a) & $> 10^{12}$ (b) \\
pulsars& &  \\
\noalign{\smallskip} \hline \noalign{\smallskip}
ms X-ray pulsars&  0.001 (c), 0.01 (d, e)& $ 3\times10^7$ (d, e), $3 \times 10^8$ (d) \\
& 0.03 (c), 0.07 (e)&  \\
\noalign{\smallskip} \hline \noalign{\smallskip}
Atoll sources  & 0.01 - 0.9 (f)  & $ 3\times 10^7$ (g)\\
\noalign{\smallskip} \hline \noalign{\smallskip}
Z  sources& 0.5 - 1 (f)     &$10^{7-8}$ (h), $3\times 10^8$ (g),       \\
\noalign{\smallskip} \hline \hline
\end{tabular}
\end{center}
[a] Chakrabarty 1996
[b] Makishima et al. 1999
[c] Chakrabarty \& Morgan 1998 
[d] Lamb \& Yu 2005 
[e] Gilfanov et al. 1998 
[f] van der Klis 1996 
[g] Zhang \& Kojima 2006 
[h] Titarchuk et al. 2001 
\end{table}

\subsection{Classical X-ray pulsars}\label{classic}

Classical X-ray pulsars\footnote{As we already stated in Section~\ref{introduction}, we are interested in accreting systems so we never refer to rotation-powered pulsars but always to accretion-powered ones.} have periods of the order of one second or more. In this sense they are also called ``slow" accretion-powered pulsars in comparison
to the millisecond X-ray pulsars (Sect.~\ref{millisec}).
Only five classical X-ray pulsars
have been found in LMXBs, whereas the vast majority are found in HMXB systems (Psaltis 2004).

\begin{figure}[t!]
\centering
\resizebox{\hsize}{!}{\includegraphics[ angle=0, scale=0.75]{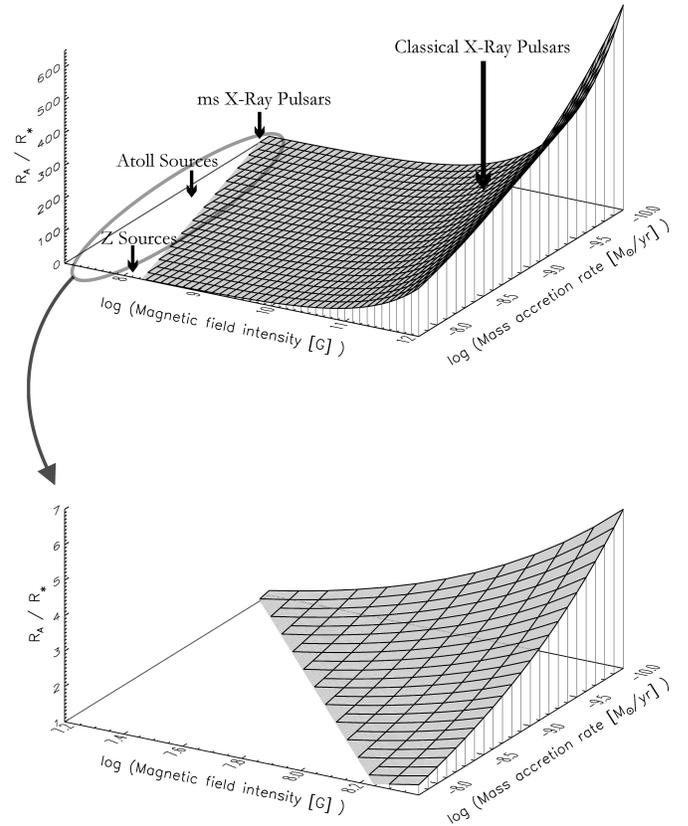}}
\caption{3-D plot of the Alfv\'en radius normalised to the stellar radius ($R_{\rm A}/R_*$). 
 The intersection between the function and the 
$R_{\rm A}/R_*= 1$ plane  indicates 
the combination of the magnetic field and the mass accretion rate values for which
plasma pressure, $P_{\rm p}$, and magnetic field pressure,
 $P_{\rm B}$, 
 balance each other 
 at the surface of the star. This ensures that the initial condition
for  jet  formation ($P_{\rm B} < P_{\rm p}$)
is fulfilled   over the whole accretion disk.
A zoom of this region is shown at the bottom of the figure.}
\label{3-D}
\end{figure}

It is clear from Fig.~\ref{3-D} that classical X-ray pulsars (known to have magnetic fields of $10^{12}$G, see Table~\ref{mdot}) have  $R_A/R_*>>1$ for any value of the mass accretion rate, even for the Eddington critical rate.
The stellar field is therefore dynamically dominant. 
In this case the plasma is forced to
move along the magnetic field lines (as shown on the right-hand side of Fig.~\ref{micro+pulsar}) and
converges onto the magnetic poles of the neutron star. There, it releases
its energy, creating two X-ray emitting caps (Psaltis 2004). In the case of a misalignment between
the rotation and the  magnetic axis, pulses are produced.

Since $R_A/R_*>>1$,
the formation of jets in classical X-ray pulsars is excluded for any accretion rate.
This  agrees with the observations:
a deep search for radio emission from X-ray pulsars was carried out
by Fender et al. (1997) and none of the pulsar candidates was detected in that wavelength.
The lack of radio emission is  discussed statistically
by Fender et al. (1997). They conclude that X-ray pulsations and radio emission
from X-ray binaries are strongly anti-correlated, which is in clear agreement
with our result.
More recent observations (Fender \& Hendry 2000; Migliari \& Fender 2006) have confirmed that  
none of the high-magnetic-field X-ray pulsars is a source of synchrotron radio emission.

\subsection{Atoll and Z-sources}\label{AandZ}

Following their timing and spectral properties (Hasinger \& van der Klis 1989), LMXBs with neutron stars have been divided into two subclasses called Atoll-type (the largest subclass) and Z-type. Both subclasses undergo different spectral states with different accretion rate values. In the case of Atoll-sources, there are two possible states, island and banana, whereas for the Z-type there are three, the horizontal-, the flaring-, and the normal-branch  (see Fig. 1 in Migliari \& Fender 2006).

Some Atoll have been detected in radio (Fender \& Hendry 2000; Rupen et al. 2005) and recently evidence of a jet has been found in some of them (Migliari et al. 2006; Russell et al. 2007). Various Z sources have also been detected in radio (Migliari \& Fender 2006). Two of them are in fact known to be microquasars (see Table~\ref{microquasars}): \object{Circinus X-1} (Fender et al. 2004b) and \object{Scorpius X-1} (Fomalont 2001). We can see from Fig.~\ref{3-D} (bottom part) that the basic condition $R_A/R_* = 1$ is satisfied if $B \leq 10^{8.2}$ G. Our upper limit agrees in fact
with the estimate of Titarchuk et al. (2001) on the magnetic field of the microquasar Scorpius X-1. They determined the value of $B$ from magnetoacoustic oscillations in kHz QPO reaching a strength of $10^{7-8}$ G on the surface of the neutron star. 

\subsection{Millisecond X-ray pulsars} \label{millisec}

Millisecond X-ray pulsars have a weak magnetic field  $B \sim 10^8 $G (Table~\ref{mdot}) together with their main characteristic of being a rapidly spinning neutron star. Very few of them have been detected up to now and they are all in the class of LMXBs.
As one can derive from the values given in Table~\ref{mdot}, millisecond X-ray pulsars
are extreme Atoll sources. The prolonged and sustained accretion of matter
on the neutron star from the long-living companion, carrying angular momentum,
is thought to be responsible for the spin up to a millisecond rotation.
Less clear is the cause for the decay of their $B$ (Cumming et al.  2001; Chakrabarty 2005; Psaltis 2004).

As shown  in Fig.~\ref{3-D},
the obstacle for  jet production in millisecond X-ray pulsars
is their low accretion rate.
For their average  $B\sim 10^8$G, as  assumed in the literature,
the basic condition $R_A/R_* = 1$ would  only be fulfilled for   
accretion rates $\dot M \geq 6\times 10^{-9} \rm{M}_{\odot}~\rm{yr}^{-1}$,
whereas  the maximum observed accretion rate is nearly one order of magnitude lower, i.e  
 $\dot M \leq 7 \times 10^{-10} \rm{M}_{\odot}~\rm{yr}^{-1}$ (Table~\ref{mdot}).
On the contrary, if $\dot M$ is in the range of $3-7 \times 10^{-10}\rm{M}_{\odot}~\rm{yr}^{-1}$,
 the basic condition $R_A/R_* = 1$ would be fulfilled
for $B=10^{7.5}$ G, which is compatible with certain observational values of the magnetic field strength (see Table~\ref{mdot}).

In fact, in  the  accreting millisecond X-ray pulsar \object{SAX J1808.4-3658}, 
the long-term mean mass transfer rate is $\dot M \simeq 1 \times   10^{-11}\rm{M}_{\odot}~\rm{yr}^{-1}$
(Chakrabarty \& Morgan 1998).
During bright states, peak values of  $\dot M \sim 3-7 \times 10^{-10}\rm{M}_{\odot}~\rm{yr}^{-1}$
(Chakrabarty \& Morgan 1998; Gilfanov et al. 1998) were measured and the   
upper limit of the magnetic field strength was found to be a  few times $10^7$ G (Gilfanov et al. 1998).
It is important to point out that in this source hints of a radio jet have been found.
There are two accreting millisecond X-ray pulsars,
SAX J1808.4-3658 (Gaensler et al.  1999) and \object{IGR J00291+5934} (Pooley 2004),
that have shown transient radio emission related to X-ray outbursts (see also  Russell et al. 2007).
Especially interesting is  that the
size of the radio emitting region of SAX J1808.4-3658 
is much larger than the separation of the binary system, which is
expected in the case of expanding  material  ejected from a system
(Gaensler et al. 1999; Migliari \& Fender 2006).
In other words, SAX J1808.4-3658, which normally behaves like a pulsar (right-hand side of Fig.~\ref{micro+pulsar}), could switch to a microquasar state at a  maximum accretion rate (left-hand side of Fig.~\ref{micro+pulsar}).  
While  future high-resolution radio observations can probe or rule out the presence of
a radio jet in millisecond X-ray pulsars, at the moment, theory and observations seem to give positive indications of it.

\section {Black hole X-Ray binaries and active galactic nuclei} \label{BHXRB}

\begin{figure}[t!]
\centering
\resizebox{\hsize}{!}{\includegraphics[ angle=0, scale=0.75]{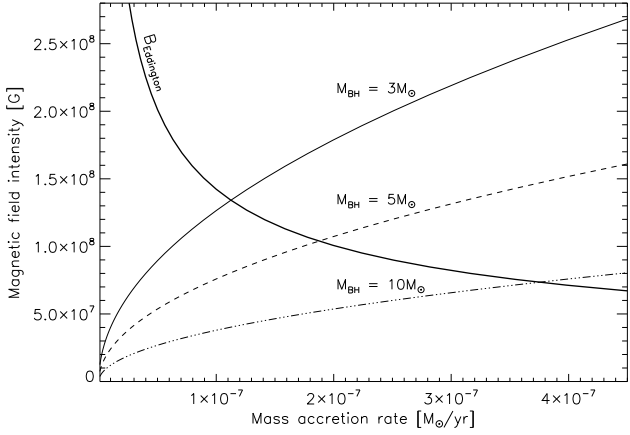}}
\caption{Schwarzschild-BH XRBs: imposing the basic condition $R_{\rm A}/R_{\rm LSO}= 1$ we obtain a relation between the magnetic field strength at the last stable orbit (LSO = $6GM_{\bullet}/c^2$) and the mass accretion rate for different values of the BH mass.  
Radiative efficiency for a Schwarzschild BH is $\eta = 0.06$.  
The  $B_{\rm Eddington}$ curve is the  upper limit for the magnetic field strength that corresponds to a mass accretion rate equal to the Eddington critical rate.
}
\label{BHXRB_Schw}
\end{figure}

To determine the so-called white area (see Section 3) where the combination of the magnetic field strength and the mass accretion rate values is such that the NS XRBs can undergo a microquasar phase, it was fundamental to have observational data of these two magnitudes, which limited their possible range of values.
 This is instead 
not possible  for  BH XRBs,  in which case
 we can  still  find out the upper limit for the magnetic field at the Eddington mass accretion rate.
 
For this purpose we equate the magnetic field pressure to the plasma pressure at the last stable orbit,\footnote{A similar procedure was used by Cheng \& Zhang (2000)  to determine the final value (i.e. for $R_A=R_*$) of the decayed magnetic field in an accreting NS.} and in this way our basic condition for the case of BH XRBs is given by $R_{\rm A}/R_{\rm LSO} = 1$. Using Eq.~\ref{ar}, where we replace $R_*$ by $R_{\rm LSO}$, we get the magnetic field strength as a function of the mass accretion rate and the BH mass. We first consider the case of a Schwarzschild BH where the LSO = $6 r_{\rm g} = 6GM_{\bullet}/c^2$:

\begin{figure}[t!]
\centering
\resizebox{\hsize}{!}{\includegraphics[ angle=0, scale=0.75]{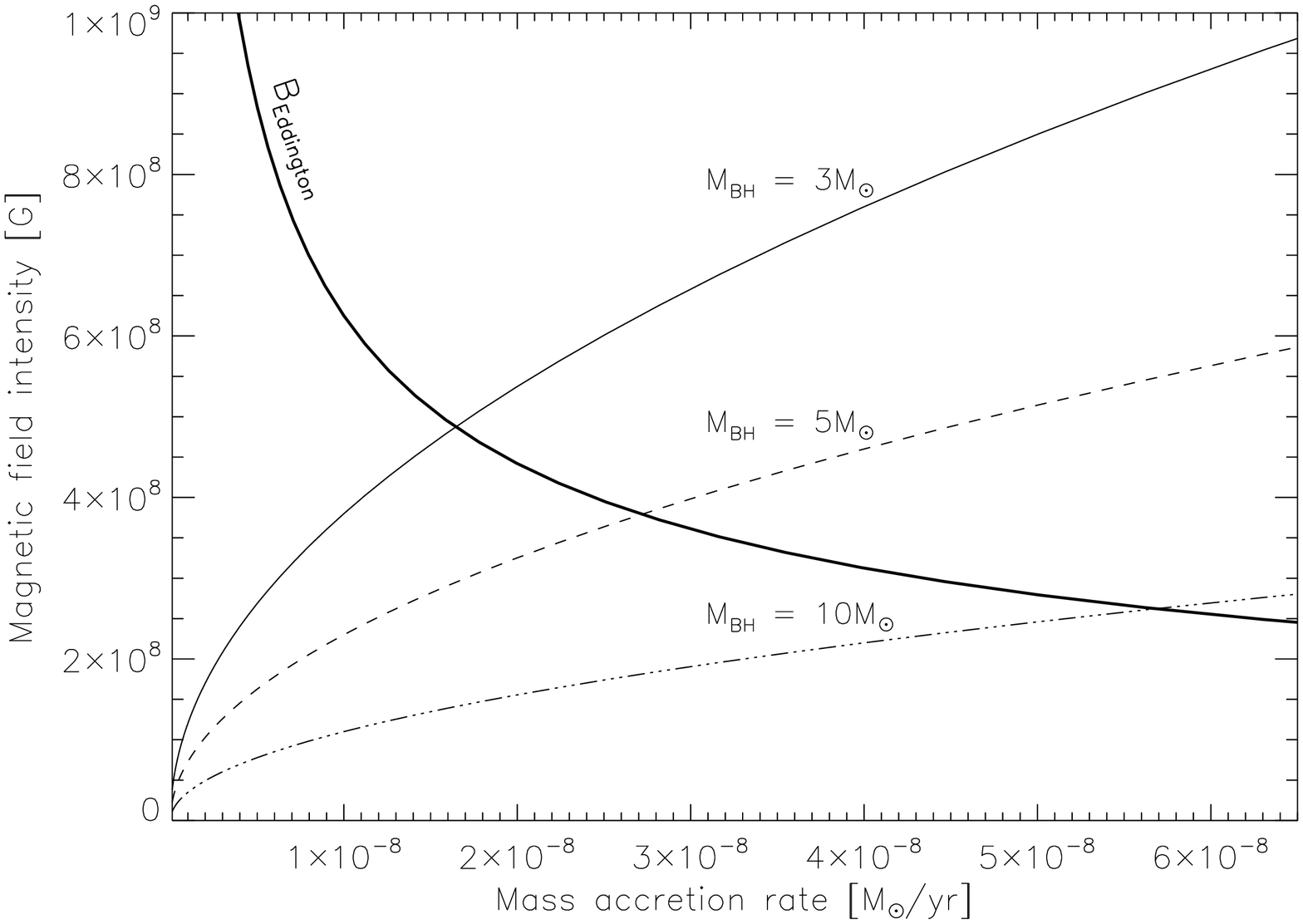}}
\caption{Kerr-BH XRBs: imposing the basic condition $R_{\rm A}/R_{\rm LSO}= 1$ we obtain a relation between the magnetic field strength at the innermost possible stable orbit (LSO = $GM_{\bullet}/c^2$) and the mass accretion rate for different values of the BH mass. Radiative efficiency for a Kerr-BH is  $\eta = 0.4$.
 The $B_{\rm Eddington}$ curve is the same  as in Fig. \ref{BHXRB_Schw}.
}
\label{BHXRB_Kerr}
\end{figure}

\begin{eqnarray}
\lefteqn{ 
B_{\rm S}= \left(3 \over c^2 \right)^{-5/4} \left( {2G}\right)^{-1}~ \left(\frac{\dot M}{M_{\bullet}^2}\right)^{1/2}{} }\nonumber\\
& & {} \simeq 1.2 ~ 10^8 \left(\frac{ M_{\odot}}{M_{\bullet}}\right)\left({\dot M\over {10^{-8}~ {\rm{M}_{\odot}\over yr}}}\right)^{1/2} \rm G.
\label{B}
\end{eqnarray}

\noindent This relation becomes

\beq
B_{\rm K}\simeq 1.1 ~ 10^9 \left(\frac{ M_{\odot}}{M_{\bullet}}\right)\left({\dot M\over {10^{-8}~ {\rm{M}_{\odot}\over yr}}}\right)^{1/2} \rm G,
\label{B_Kerr}
\eneq

\noindent when considering an extreme  Kerr BH where the innermost possible stable orbit is instead LSO = $GM_{\bullet}/c^2$.

In Figs.~\ref{BHXRB_Schw} and \ref{BHXRB_Kerr}, we show the result of evaluating Eqs.~\ref{B} and \ref{B_Kerr}, respectively, for different values of stellar-mass BHs. 
Using Eqs.~\ref{B} and \ref{B_Kerr} and the relation

\beq
\dot M_{Edd} \approx \frac{2 ~ 10^{-9}}{\eta} \frac{M}{M_{\odot}}\frac{M_{\odot}}{yr},
\label{accedd}
\eneq

\noindent we obtain the upper limit for the magnetic field strength that corresponds to a mass accretion rate equal to the Eddington critical rate (``$B_{\rm Eddington}$" curve in Figs.~\ref{BHXRB_Schw} and \ref{BHXRB_Kerr}), where $\eta$ is the radiative efficiency with values $\eta = 0.06$ for a Schwarzschild BH and $\eta = 0.4$ for a Kerr BH.
It can be seen that the upper limit of the magnetic field strength for the whole range of stellar-mass BHs, for the maximum possible accretion rate, is $B_{\rm S} \simlt 1.35 \times 10^8$ G and $B_{\rm K}  \simlt 5 \times 10^8$ G in the case of a Schwarzschild or Herr BH, respectively. 

\begin{figure}[t]
\centering
\resizebox{\hsize}{!}{\includegraphics[angle=0, scale=0.75]{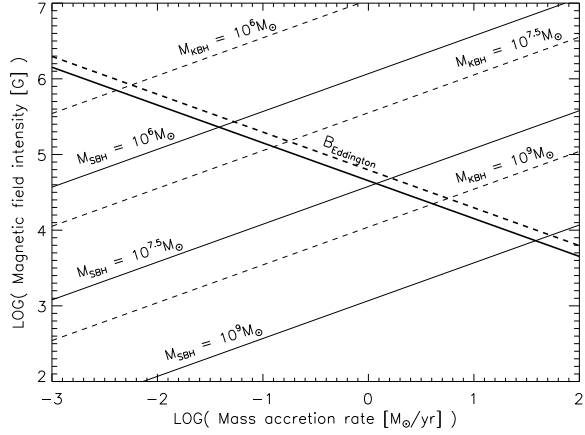}}
\caption{Supermassive BHs: using Eqs. \ref{B} and \ref{B_Kerr} we obtain by standard scaling a relation between the magnetic field strength at the last stable orbit and the mass accretion rate for different values of the mass of supermassive BHs. The cases of Schwarzschild and Kerr BHs are considered here with solid and dashed lines respectively.
The  $B_{\rm Eddington}$ curve is the same as in Fig. \ref{BHXRB_Schw}.
}
\label{agn}
\end{figure}

Equations \ref{B} and \ref{B_Kerr} shows the straightforward dependency of the magnetic field strength with the mass of the BH allowing us to establish its upper limit for the jet formation in the case of supermassive BHs as well (see Fig. \ref{agn}). 

Taking the maximum Eddington critical rate  into account, we can see that the magnetic field strength at the last stable orbit for jets to be formed has an upper
limit of $B \simlt 10^{5.4}$~G for  Schwarzschild BHs and $B \simlt 10^{5.9}$~G for the  Kerr BHs.

It is worth noting that
 we get  $B \simlt 10^{4.3}$~G in the specific case of a  supermassive Schwarzschild BH of $10^8 ~ \rm M_{\odot}$.
For a BH of the same mass, Blandford \& Payne (1982) established  $B \simlt 10^{4}$ G at  10 $r_{\rm g}$. 
Scaling our value, which is relative to LSO=6$r_{\rm g}$, to  10$r_{\rm g}$,
 we get  $B \simlt 10^{4.0}$ G in complete agreement with the results of  Blandford \& Payne (1982).

\section {Magnetic field decay} 
Measurements of surface magnetic field strengths
 by cyclotron resonance effects
in  a dozen of X-ray classical pulsars 
show that $B$ is  tightly concentrated over a narrow range of (1 - 4) $\times 10^{12}$ G (Makishima et al. 1999).
Under the condition of magnetic flux conservation, a neutron star with a magnetic field $B\sim 10^{12}$ G will result from a progenitor star with radius $R_*\sim 10^{11}$ cm and magnetic field $B\sim 10^{2}$ G.
To achieve the basic conditions for forming a jet, the
field must decay to B$=10^{7-8}$G (see previous sections).

Analysis of pulsars data have indicated that a magnetic field decays 4 orders of magnitude by Ohmic dissipation in a timescale longer than
$10^9$ yr (Konar \& Bhattacharya 2001 and references therein). Therefore this kind of magnetic field decay process excludes the possibility of a NS-HMXB 
evolving into a microquasar phase since this decay is longer than the lifetime of the high-mass companion star, i.e. $\simlt 10^7$yr for $M_* \sim 10 \rm M_{\odot}$. In this case then, the only possible accretor would be a BH.

However faster decays of the magnetic field can occur with the high-accretion-induced crust screening process (Zhang 1998). The case of Circinus X-1 is very interesting
 in this context. Circinus is a NS-LMXB\footnote{The accretor of this system is clearly a type I X-ray bursts NS.} with jet, i.e. a microquasar (Table~\ref{microquasars}), and is so young that its orbit has not yet had time to become circular (circularization time $\sim 10^5$yr, Ransom et al. 2005). In order to have already reached the magnetic field value to fulfill the  basic condition
derived here for jet formation, i.e. $B\sim10^{7-8}$G, the magnetic field decay time due to screening has to have been shorter than $10^5$yr. In fact, Romani (1995) has deduced a characteristic timescale for the initial field decay in the range of $10^4 \rm yr \leq t \leq 10^6 \rm yr$. In this context, when the magnetic field decays is due to screening, a NS microquasar will rotate
very fast because of the large angular momentum transfer due to the heavy accretion.

\section {Discussion and conclusions} 

We have analysed the initial condition for an XRB to undergo a microquasar phase, i.e. to generate jets. The relation between the $P_B$ and the $P_p$ was studied using the $R_{\rm A}/R_*$ or $R_{\rm A}/R_{\rm LSO}$ ratios. The basic condition for jet formation, $R_{\rm A}/R_* = 1$ or $R_{\rm A}/R_{\rm LSO}=1$, led us to quantify an upper limit for the magnetic field strength as a function of the mass accretion rate. In this context, we studied each of the possible accreting XRB systems with neutron stars or black holes as the compact objects and we reached the following results:

\begin{enumerate}
\item
The association of a classical X-ray pulsar (i.e. $B \sim 10^{12}$ G) with jets is excluded even if they accrete at the Eddington critical rate, in agreement with the systematic search of radio emission in this kind of sources
 with so far negative results (Fender et al. 1997, Fender \& Hendry 2000; Migliari \& Fender 2006).
\item
It is known that Z-sources, ``low" magnetic field neutron stars accreting at the Eddington critical rate, may develop jets. In this work we have 
quantified the magnetic field strength to be $B \simlt 10^{8.2}$ G in order to make possible the generation of jets in this kind of sources. This upper limit fits the observational estimation of Titarchuk et al. (2001) for Scorpius X-1. 
\item
Atoll-sources are potential sources for generating jets  if $B \simlt 10^{7.7}$~G. In fact evidence  of jets in these sources has been  found in Migliari et al. (2006) and Russell et al. (2007).
\item
It is not ruled out that a millisecond X-ray pulsar could develop jets, at least for those sources
where  $B \simlt 10^{7.5}$ G. In this case
the millisecond X-ray pulsar (right-hand side of Fig.~\ref{micro+pulsar}), could
 switch to a microquasar phase during maximum accretion rate (left-hand side of Fig.~\ref{micro+pulsar}).
The millisecond source SAX J1808.4-3658 with such a low $B$
shows in fact hints of a radio jet.
\item
In the case of BH XRBs, the upper limit of the magnetic field strength for the whole range of stellar-mass BHs, taking an Eddington mass accretion rate  into account, is $B \simlt 1.35 \times 10^8$ G and $B \simlt 5 \times 10^8$ G for a Schwarzschild and a Kerr black hole, respectively.
In this context we also studied the magnetic field strength upper limit for the jet generation in AGNss and found $B \simlt 10^{5.4-5.9}$ G, again taking the two spin extremes  into account.
\end{enumerate}

It is worth noting in Fig.~\ref{chart} that two cycles (A and B) are indeed strongly linked. When the basic condition is reached in cycle A, a state transition occurs and 
the system enters  cycle B, i.e. a jet is formed. 
This excludes the presence of a jet in the soft X-ray state, in which the system is found
 during cycle A. 
The lack of a jet  instead
does not exclude the presence of radio emission during the soft state. This issue is in fact addressed by Laor \& Behar (2007) for the case of the radio emission in radio-quiet AGNs (i.e. without jets) that is shown to originate in a magnetically active corona above the accretion disk. 

This analysis of the basic condition for jet formation 
has as well some important implications.  
A decay  
  in  the magnetic field  due only to Ohmic dissipation 
 implies the presence of a BH as the compact object in a microquasar-HMXB  because of the long
timescales of this process. Only in the case of 
high-accretion-induced crust screening process the 
timescales can be as short as $\rm t \simlt 10^5$yr and the issue of the nature of the compact
object remains open.

Finally, the last implication is related to the possibility of jets in millisecond X-ray pulsars.
One of the major open issues concerning millisecond X-ray pulsars is the absence
(and possible non-existence) of sub-millisecond X-ray pulsars. The spin distribution sharply
cuts off well before the strict upper limit on the NSs spin rate that is given by the centrifugal breakup limit (0.3 ms depending on the NS equation of state). The physics setting that limit
is unclear (Chakrabarty 2005).
If the jet hypothesis is finally proved, then the jet 
might be the suitable agent of angular momentum sink,
as  in the bipolar outflows from young stellar objects.
The  transport rate of angular momentum by the jet can be two thirds or more of the estimated rate transported through the relevant portion of the disk (Woitas et al. 2005).

\begin{acknowledgements}
We are grateful to Peter L. Biermann and Karl Menten for very useful comments on the manuscript. The authors are grateful to an anonymous referee, whose valuable suggestions and comments helped to improve the paper.

\end{acknowledgements}

\end{document}